\documentclass[aps,pre,twocolumn,amsmath,amssymb,showpacs,amsfonts]{revtex4}
\usepackage{epsfig}
\usepackage{graphicx}
\usepackage{dcolumn}
\usepackage{bm}

\begin{document}

\title{The bound on viscosity and the generalized second law of thermodynamics}

\author{Itzhak Fouxon}
\author{Gerold Betschart}
\author{Jacob D. Bekenstein}

\affiliation{Racah Institute of Physics, Hebrew University of Jerusalem,
Jerusalem 91904, Israel}
\date{\today }

\begin{abstract}

We describe a new paradox  for ideal fluids. It arises in
the accretion of an \textit{ideal} fluid onto a black hole, where, under suitable boundary conditions,  the flow can violate
the generalized second law of thermodynamics. The paradox indicates that there is in fact a lower bound to the correlation length of any \textit{real} fluid, the value of which is determined by the thermodynamic properties of that fluid.   We observe that the universal bound on entropy, itself suggested by the generalized second law, puts a lower bound on the correlation length of any fluid in terms of its specific entropy.  With the help of a new, efficient estimate for the viscosity of liquids, we argue that this also means that viscosity is bounded from below in a way reminiscent of the conjectured Kovtun-Son-Starinets lower bound on the ratio of viscosity to entropy density.   We conclude that much light may be shed on the Kovtun-Son-Starinets bound by suitable arguments based on the generalized second law.

\end{abstract}


\maketitle

\section{Introduction}

The use of general principles to investigate systems whose microscopic makeup is unclear can be very rewarding. Sometimes this method gives information on a whole class of systems. Among such principles, thermodynamics, believed to be of universal applicability, stands out. An instructive example of its use is the application of the second law of thermodynamics to the problem of viscous flow, for which it permits the inference that the two viscosity coefficients must be positive without need to resort to microscopic expressions for the latter~\cite{Landau6}.

In this work we use the generalization of the second law of thermodynamics holding in the presence of black holes (the  generalized second law---GSL)
to reach further conclusions about the shear viscosity coefficient of an arbitrary fluid.  We do this by describing a new paradox
for ideal fluid flow in the presence of a black hole.  This indicates that the correlation length of a real fluid cannot be arbitrarily small.  By implication the energy-momentum tensors describing systems which display macroscopic fluid behavior must be subject to a restriction: the shear viscosity, a function of the thermodynamic state of the system, cannot be arbitrarily small.  Thereby the GSL opens an alternative macroscopic approach to the recently proposed lower bound on viscosity~\cite{KSS1,KSS2}.

The GSL is a unique law of physics that bridges thermodynamics and
gravity.  It is rooted in the understanding that a black hole, basically a pure gravity entity, is endowed with well defined entropy~\cite{JB1,JB2,Hawking1} proportional to its surface area.  The GSL~\cite{JB1,JB22} then claims that the sum of the
entropy of all black holes and the total ordinary entropy in the black holes' exterior never decreases.  While this formulation is reminiscent of the ordinary second law, the GSL is exceptional in that it relates ordinary entropy---a rather elusive object from the mechanical viewpoint---and the surface area of the
black hole (formally  the area of its horizon) whose evolution is quite mechanical in nature.  From this point of view it is little surprise that the GSL has provided unexpected information on entropy.

An example is the upper bound on the entropy of weakly self-gravitating thermodynamic systems (the universal bound on entropy---UBE~\cite{JB3,JB4}).  While in particular cases the bound can be verified directly,
it is the GSL which really makes understanding of the bound in generic situations easy. However, the GSL gives more than just a simpler way
to see some results derivable by other means.  At the microscopic level the GSL represents a piece of the yet to be established theory of quantum gravity. In particular, this law permits, in principle, to draw conclusions that from the microscopic viewpoint would only be derivable from a fundamental theory combining
quantum mechanics with  gravity. For example, the GSL gives an indication of the number of particle species in nature~\cite{JB4,JB5}.  In another example, the GSL reveals an \textit{a priori}  bound on the strength  of the electromagnetic interaction~\cite{Davies}.

Let us now describe the ideal fluid paradox revealed by invoking the GSL. We assume the existence
of physical fluids with arbitrarily small shear viscosity $\eta$ at fixed values of some two thermodynamic variables,
say the entropy and the energy densities, $s$ and $\rho$, respectively.  (In this work we
consider only simple fluids, so the values of $s$ and $\rho$ completely determine the thermodynamic state
of the system.)  Here ``physical" means, among other things, that the fluid satisfies
the GSL; this could,  in principle, be checked by considering the fluid at the microscopic level, but we shall not go into such detail.

The assumption that the fluid has arbitrarily small viscosity allows us to describe its flow as
ideal fluid flow, possibly with shocks (the zero viscosity limit of a given flow can be non-trivial; see Refs.~\cite{Frisch,Landau6}).  It turns out, as we shall show, that for sufficiently slow accretion of the fluid onto
the black hole the overall entropy decreases and the GSL is violated.   The realizability, in principle, of such slow accretion flows will be demonstrated, so that the small viscosity assumption engenders a paradox.

Of course, ideal fluid paradoxa exist already in non-relativistic physics, for example, the famous d'Alembert paradox.  This maintains
that an ideal fluid with no boundaries exerts no force whatsoever on a body moving through it with constant velocity.  In particular,
there is no lift force, so swimming or flying would be impossible in such a fluid. Thus the established fact that swimming is possible in any fluid implies that  every real fluid must have nonzero viscosity.  There is however an essential difference between the d'Alembert paradox and the paradox described in this work. D'Alembert's paradox says nothing about the actual magnitude of $\eta$.
In contrast, our paradox implies, as we shall see,  the existence of a lower bound on $\eta$ for given $s$ and $\rho$.

This conclusion is in concord with the recent ingenious conjecture of Kovtun, Starinets and Son (KSS)~ \cite{KSS1,KSS2} (see also Ref.~\cite{SO}). Based on holographic calculations of the viscosity coefficient for certain strongly coupled quantum field theories with gravity duals, they suggested that the viscosity $\eta$ of a general, possibly non-relativistic, fluid is subject to the universal
restriction
\begin{equation} \label{KSS}
\frac{\eta}{s}\geq \frac{\hbar}{4\pi}\ .
\end{equation}
Currently this bound is considered a conjecture well supported for a certain
class of field theories---see the detailed discussion in Ref.~\cite{ChermanCohen} and the references therein.
In the sequel we discuss the relation between the ideal fluid paradox presented in the next section and the KSS bound.  We also argue that a frequent objection to the validity of the KSS bound is likely to be ruled out by the GSL.

Our paper is structured as follows. In Sec.~\ref{paradx} we show explicitly that the slow accretion of a truly ideal fluid onto a Schwarzschild  black hole leads to a contradiction with the GSL.  One escapes from the paradox by recognizing that every fluid must have a nonvanishing correlation length which restricts the range of applicability of the ideal fluid paradigm.  In Sec.~\ref{KSSp} we obtain a lower bound on the correlation length and a generic estimate of the viscosity of real fluids, which together bound the viscosity to entropy density ratio from below.   Although this is not yet the KSS bound, we consider there the connection between it  and the UBE, and argue that the GSL provides a natural frame for elucidation of the origin of the former.  In the Sec.~\ref{summary} we summarize our results and arguments. The realizability, in principle, of the slow accretion flow assumed in Sec.~\ref{paradx} is demonstrated  in the Appendix.

Unless otherwise stated, we work in units with $c=\hbar=k=1$, where $c$ is speed of light and $k$ is Boltzmann's constant.  Our metric signature is $(-,+,+,+)$.

\section{Ideal fluid paradox}
\label{paradx}

In the present section we consider ideal fluid flow into a  spherical black hole. For some flows,  we demonstrate that  it is possible for the GSL to be violated so that the total entropy of the system decreases. It follows that the assumption of a perfect continuum down to an arbitrarily small
scale is not consistent with the GSL.

\subsection{Entropy balance in accretion onto a black hole}
\label{balance}

Consider a flow, not necessarily spherically symmetric, in which fluid is absorbed by the black hole.   The rate of change of the total entropy $S$ of the system is the sum of the rate of change of the entropy of the black hole exterior, $S_{ext}$,
and that of the black hole entropy $S_{H}$:
\begin{equation}
\frac{dS}{dt}=\frac{dS_{ext}}{dt}+\frac{dS_{H}}{dt}. \label{total}
\end{equation}
Here and below we use Schwarzschild coordinates $(x^0,x^1,x^2,x^3)=(t, r, \theta, \phi)$, with $r\geq r_H$ where $r_H$ is the Schwarzschild radius.

We first calculate $dS_{ext}/dt$.  Let the fluid's proper entropy density be $s$.  Since the fluid is assumed ideal, there is no dissipative contribution to the entropy current density which is thus purely convective, and must take the form $s U^\mu$, where $U^{\mu}$ is the fluid four-velocity.   The fluid can
carry entropy into the hole leading to a decrease of $S_{ext}$. The explicit expression
for this comes from the entropy balance equation~\cite{Landau6,Landau2,Weinberg},
\begin{equation}
\partial_\mu(\sqrt{-g}sU^{\mu})=0,
\end{equation}
where $g=-r^4\sin^2 \theta$ stands for the determinant
of the Schwarzschild metric $g_{\mu\nu}$.

$dS_{ext}/dt$ of  the black hole exterior ($r>r_H$) is thus
\begin{eqnarray}
&&\frac{d}{dt}\int_{r_H} s U^0\sqrt{-g}dr d\theta d\phi=
-\int_{r_H}dr d\theta d\phi\,  \partial_r(s U^{r} \sqrt{-g})\nonumber \\
&&=\int_{r=r_H}  s U^{r} \sqrt{-g} d\theta d\phi.
\label{exterior}
\end{eqnarray}
We have not included a contribution from the outer boundary of the domain of integration because we intend to specialize to stationary flows.
In any such situation the entropy flow into the hole per unit $t$-time is given by the r.h.s. of Eq.~(\ref{exterior}).
The expression is non-positive because $U^r\leq 0$ for infalling matter.
We have assumed that the flow is differentiable, and that it contains no shocks; otherwise
there exists an additional contribution to $dS_{ext}/dt$ associated with the entropy generation
in the shocks~\cite{Landau6}.

Let us now consider the second term of the r.h.s. in Eq.~(\ref{total}). The absorption of the fluid by the
black hole increases the latter's mass $M$, producing an increase in the black
hole entropy given through energy conservation by
\begin{equation}
T_{H}\frac{dS_{H}}{dt}=\frac{dM}{dt}=-\frac{dE_{ext}}{dt},
\end{equation}
where $T_{H}$ is the black hole temperature, and $M$ is its mass. We shall now write down the flux
of energy into the black hole.  Let $\bm \xi^t=(1, 0, 0, 0)$ be the Killing vector associated with the stationarity of the black
hole.  Then the energy-momentum tensor of the fluid, $T^{\mu\nu}$, must obey
\begin{equation}
(\bm\xi^t_{\nu}T^{\mu\nu})_{; \mu}=0,
\end{equation}
or equivalently
\begin{equation}
\partial_\mu(\sqrt{-g}T^{\mu}_0)=0.
\end{equation}
Since $T^0_0\sqrt{-g}$ is \textit{minus} the energy density of the fluid,
\begin{eqnarray}
T_{H}\frac{dS_{H}}{dt}=-\frac{dE_{ext}}{dt}&=&\frac{d}{dt} \int_{r_H} T^{0}_{0}  \sqrt{-g}dr d\theta d\phi
\nonumber\\
=-\int_{r_H}dr d\theta d\phi\,\partial_r
(T^{r}_{0}  \sqrt{-g})&=&\int_{r=r_H} T^{r}_{0}  \sqrt{-g} d\theta d\phi.
\end{eqnarray}

The energy-momentum tensor of the ideal fluid is given by
\begin{equation}
\label{Tmunu}
T^{\mu\nu}=(\rho+p)U^{\mu}U^{\nu}+pg^{\mu\nu}, \end{equation}
where $\rho$ is the energy density and $p$ is the pressure in the comoving frame.
Consider now the normalization condition $U^{\mu}U_{\mu}=-1$, written as
\begin{equation}
U_0^2-(U^r)^2=g_{rr}^{-1}[1+g_{\theta\theta}(U^{\theta})^2+g_{\phi\phi}(U^{\phi})^2].
\end{equation}
Since $\sqrt{g_{\theta\theta}}\,U^{\theta}$ and $\sqrt{g_{\phi\phi}}\,U^{\phi}$ are physical velocity components, they should be bounded at $r=r_H$.  Hence since $U^\mu$ is future and inwardly pointed, and $g_{rr}\rightarrow\infty$ as $r\rightarrow r_H$, we can infer from the last equation that
$U_0=U^r$ at the horizon . Combining all the above we find
\begin{equation}
\frac{dS_{H}}{dt}=\frac{1}{T_H}\int_{r=r_H}(\rho+p)(U^r)^2
\sqrt{-g} d\theta d\phi.
\label{black}
\end{equation}
Thus $dS_{H}/dt>0$ in harmony with Hawking's area theorem~\cite{hawk}.

We observe from Eqs.~(\ref{exterior}) and~(\ref{black}) that while the (negative) rate of change of $S_{ext}$  is proportional to the first power of $U^r$, the (positive) rate of change of the black
hole entropy is proportional to the \textit{second} power of $U^r$.   Thus for sufficiently small $U^r(r_H)$, the total entropy of the system will \textit{decrease}, in violation of the GSL. Explicitly we have
\begin{equation} \label{entr2}
\frac{dS}{dt}=-\int_{r=r_H} |U^r|
\sqrt{-g} d\theta d\phi\left[s-\frac{|U^r| (\rho+p)}{T_{H}}\right],
\end{equation}
where we stress that at the horizon $U^r$ is never positive.
We observe that when the accretion
velocity $U_{ac}$ (the suitable mean value of $|U^r|$ over the horizon)
obeys
\begin{equation}
U_{ac}< \frac{s T_H}{\rho+p}, \label{ineq0}
\end{equation}
the total entropy decreases and the GSL is broken.  We now proceed to search for such flows.  We first reconsider known explicit solutions.

\subsection{Validity of the GSL for the Bondi flow}

The above formulae apply for a generic, not necessarily spherically symmetric, accretion flow onto the black hole;  in particular it may have non-vanishing $U^{\theta}$ and $U^{\phi}$.
As an example, consider the well known Bondi flow (see Ref.~\cite{Shapiro}
and references therein). Bondi flow starts from rest at infinity and is spherically symmetric.  For a polytropic equation of state,
\begin{equation}
\label{poly}
p=K n^{\Gamma},
\end{equation}
where $\Gamma$ is the adiabatic exponent, the accretion velocity is given by $U_{ac}\approx 1$ for
$\Gamma\neq 5/3$ and $U_{ac}\approx 0.782$ for $\Gamma=5/3$
\cite{Shapiro}. It follows that  Bondi flow will obey the GSL if the following condition holds:
\begin{equation}
\label{Bondi}
\frac{s}{\rho+p}\leq
\begin{cases}
       4\pi r_H, \ \ &\Gamma\neq 5/3,\\
       (0.782)\times 4\pi r_H,  \ \ &\Gamma=5/3.
\end{cases}
\end{equation}
We have used the usual expression $T_H=(4\pi r_H)^{-1}$ for the Schwarzschild black hole.

The above conditions would seem to obtain for any fluid with positive pressure thanks to the UBE~\cite{JB3}. This bound states that the entropy $S_b$ of a generic, weakly self-gravitating, thermodynamic system satisfies
\begin{equation}\label{Bekensteinentropy}
\frac{S_b}{E_b}< 2\pi R_b,
\end{equation}
where $E_b$ is the body's total energy,  while $R_b$ is its linear size.  Our experience is that the inequality here is usually a strong one.

Now the minimal size of a parcel of fluid consistent with the fluid description is its correlation length $l$.
For a gas, $l$  is the mean free path, while for a liquid it is typically the intermolecular distance.
Applied to such a parcel the UBE tells us that (see the next Section for more details)
\begin{equation}
s/\rho< 2\pi l, \label{Bekensteinforfluid}
\end{equation}
where we have passed from the total entropy and energy to their densities by dividing by the parcel's proper volume. Again, this bound will in most cases be a strong inequality.  Since the continuum description down to accretion at the black hole makes sense only if $l\ll r_H$,
we may conclude that the inequalities~(\ref{Bondi}) will always hold with at least one order of magnitude difference between the r.h.s. and the l.h.s. of the equations.  In the above argument we have tacitly assumed that $p>0$.  Negative pressure is often discussed in cosmology (dark energy); the functioning of the GSL in the face of dark energy is fraught with subtleties~\cite{Zhou}.

Thus Bondi flow satisfies the GSL by virtue of the UBE.  Clearly, this happens because the accretion velocity approaches the speed of light, $U_{ac}\sim 1$.  We now turn to examples of slow ideal fluid accretion flows that do violate the GSL.

\subsection{Paradox in slow accretion flow of ideal fluid}
\label{paradox}

It is clear from the previous analysis  that the GSL holds for any accretion flow of ideal fluid with $U_{ac}\sim 1$.  We expect such flows whenever spherical accretion starts from a distance $r_i$ from the black hole large compared to $r_H$ (for Bondi flow
$r_i=\infty$).  But what is $U_{ac}$ in the limit $r_i\to r_H$? We now show that $U_{ac}$ vanishes in that limit, thereby giving an example of ideal fluid flow that violates  the GSL.

We study steady, spherically symmetric accretion flow which starts at $r=r_i$ with the initial flow and sound velocities, $u_i$ and $a_i$, specified there.   We assume that the relevant initial conditions
are physically realizable; see the Appendix for the discussion of this assumption.
Following Ref.~\cite{Shapiro} we introduce $u\equiv-U^r>0$.
The continuity and the Euler equations can be written as
\begin{eqnarray}&&\label{derivs}
u'=\frac{D_1}{D},\ \ n'=-\frac{D_2}{D},
\end{eqnarray}
where a prime denotes $\partial/\partial r$, $n$ is the baryon number density, and
\begin{eqnarray}&&
D=\frac{u^2-(1-2M/r+u^2)a^2}{un}, \label{D} \\&& D_1=\frac{1}{n}\Biggl[
(1-2M/r+u^2)\frac{2a^2}{r}-\frac{M}{r^2}\Biggr],\\&& D_2=\frac{2u^2/r-M/r^2}{u},
\end{eqnarray}
with $a$ the local sound velocity. Obviously $D(r=r_H)=(u/n)(1-a^2)$ is always positive due to the causality constraint $a^2<1$ (sound velocity is
smaller than velocity of light)~\cite{Shapiro}.  Here we consider only flows with $U_{ac}>0$ so
$u\neq 0$ at the horizon.

In Bondi flow $D$ becomes negative at large $r$ which signifies that there exists the so-called sonic
radius $r_s$ such that $D(r=r_s)=0$. As is clear from Eqs.~(\ref{derivs}), where $D$ stands in the
denominator, the sonic radius is a special though regular point of the Bondi flow,
which in other situations could signify the presence a shock. By contrast, here we shall choose a range of initial parameters for the flow which ensure that $D(r=r_i)>0$, so that there is no sonic point.
We assume initial parameters for the flow satisfying
\begin{eqnarray}&&
 1-\frac{2M}{r_i}\ll 1,\ \ a_i^2\ll 1.\label{in1}\\
&&\left(1-\frac{2M}{r_i}\right)a_i^2<u_i^2\ll 1-\frac{2M}{r_i},
\label{in2}\\
\end{eqnarray}
It is easy to see from Eq.~(\ref{D}) that $D(r_i)$ is indeed positive.

We do not need to derive
the explicit form of the flow in order to find $U_{ac}$. It suffices to consider the conservation of the  Bernoulli integral along the streamlines. The relativistic version of the Bernoulli equation reads~\cite{Shapiro}
\begin{equation}\label{Bern}
\left(\frac{\rho+p}{n}\right)^2\left(1-\frac{2M}{r}+u^2\right)=\textrm{const.}
\end{equation}
Assuming for simplicity the polytropic equation of state~(\ref{poly}) one finds the following relation
\cite{Shapiro}
\begin{equation}\label{Bern1}
\frac{\rho+p}{n}=m\left[1+\frac{a^2}{\Gamma-1-a^2}\right],
\end{equation}
where $m$ is the baryon mass. The above relation allows us to rewrite the Bernoulli
equation as
\begin{equation}\label{Bern2}
\left(1+\frac{a^2}{\Gamma-1-a^2}\right)^2\left(1-\frac{2M}{r}+u^2\right)=\textrm{const.}
\end{equation}
Evaluating the above equation first at $r=r_H$ and then at $r=r_i$, equating the results,  using inequalities~(\ref{in1})-(\ref{in2}) and assuming that at the horizon $a^2$ remains  much smaller than unity (which we have verified numerically), we find that
\begin{equation}
U_{ac}= \sqrt{1-\frac{2M}{r_i}}+\cdots
 , \label{accrvel}
\end{equation}
where the ellipsis stands for subleading terms. The above expression holds for any $K$ and $\Gamma$, the latter assumed to be not too close to unity.

Actually, the use of the polytropic equation is not essential.  Using Eq.~(\ref{Bern}) directly we would find
\begin{equation}
U_{ac}=
\frac{n(r_H)\left[\rho(r_i)+p(r_i)\right]}{n(r_i)\left[\rho(r_H)+p(r_H)\right]}
\sqrt{1-\frac{2M}{r_i}}.
\end{equation}
If we may assume that there are no abrupt changes  in $(\rho+p)/n$ throughout the flow, we recover a result of form Eq.~(\ref{accrvel}).
We may conclude that an ideal fluid accretion flow which starts close to the horizon with initial parameters complying with inequalities~(\ref{in1})-(\ref{in2}) can have arbitrarily small
$U_{ac}$.  According to the criterion~(\ref{ineq0}) it will thus incur a violation of the GSL.

Of course for the above argument to be convincing, we must still demonstrate that it is possible for a flow to start near the horizon with sufficiently small velocity (and sound speed). In the Appendix we show in detail that a cord which respects fundamental physical constraints can be used to bring objects to rest arbitrarily near the horizon.  One can envision the setting up of the initial conditions for the fluid we require at $r=r_i$ by this means.

\subsection{How to remove the paradox?}
\label{remove}

The above violation of the GSL raises a paradox.   In the real world the GSL cannot be violated.  This law has been shown to follow from fundamental concepts in quantum theory and classical gravity~\cite{Sorkin}.   Can the paradox be removed without calling on new physics just for the occasion?

An obvious solution would be to call on Hawking's radiance to generate entropy that would at least compensate for the decrement of total entropy pointed out in Sec.~\ref{paradox}.  After all, in free Hawking emission the thermal radiation entropy creation rate exceeds the associated rate of decrease of the black hole entropy~\cite{JB6}.  But there exist reasons to reject this as the resolution of our problem.  The Hawking radiation is capable of preventing a violation of the GSL that would arise if high entropy radiation of the same nature, i.e., electromagnetic, were injected into a black hole with $T_H$ above the effective  radiation temperature~\cite{JB6}.  Its efficacy here is related to the principle of detailed balance in equilibrium (in fact, for incoming thermal radiation at temperature $T_H$ the Hawking radiation exactly balances the entropy decrement).  Now detailed balance refers to modes of the same physical system.  There is no detailed balance between radiation and fluid modes.  Thus if entropic radiation were injected alongside the fluid accretion, the Hawking radiation might prove incapable of  compensating for the entropy decreases of the two kinds.

The above discussion also suggests we should look for a way out of the paradox that hinges on the physics of the fluid itself.   In our discussion in Sec.~\ref{paradox} we assume that one can deposit the fluid at an arbitrarily small distance from the horizon at an arbitrary velocity.  Since the physical paradigm used is that of fluids, one should, as a matter of principle,
demand that the said distance is still larger than the correlation length $l$ at which the hydrodynamic description first becomes applicable.

To be in the Schwarzschild spacetime a small proper length $l\ll M$  away from the horizon at $r_H=2M$ corresponds to the Schwarzschild coordinate $r=2M+M\delta_l$; here $\delta_l\ll 1$ and is given explicitly by
\begin{equation}
\delta_l\approx \frac{l^2}{8M^2}\, .
\end{equation}
Let us thus substitute $2M+M\delta_l$ for $r_i$ in our expression~(\ref{accrvel}):
\begin{equation}
\label{U}
|U^r|\approx \sqrt{1-\frac{2M}{r_i}}\approx \frac{l}{4M}\,.
\end{equation}
Thus the correlation length limits the smallness of the accretion velocity $U_{ac}$.

With Eq.~(\ref{U})  the expression in square brackets in Eq.~(\ref{entr2}) takes the form
\begin{equation}
[s-2\pi l(\rho+p)]\,,
\end{equation}
where we used the expression for the black hole temperature.  In order for the GSL to be obeyed the above factor must be nonpositive.  This is actually guaranteed by the fluid version of the UBE, Eq.~(\ref{Bekensteinforfluid}), as long as the pressure $p$ is nonnegative (see the next Section for more details).
This removes the paradox.  The moral of the discussion is that a paradox arises if one relies on the continuum description of a fluid down to an arbitrarily small scale, that is if one takes the notion of ideal fluid literally.  To be rid of paradoxa one must take cognizance of the finite correlation length of any physical fluid, \textit{and} must accept that the entropy capacity of fluid matter is limited according to bound~(\ref{Bekensteinforfluid}).

\section{GSL as framework for elucidating the KSS bound}
\label{KSSp}

 The paradox uncovered in Sec.~\ref{paradox} and its resolution in Sec.~\ref{remove} clearly show that the ideal fluid paradigm is inconsistent.   In particular, the picture of a fluid as a perfect continuum is shown to be physically unacceptable: the fluid in question must have a nonvanishing correlation length. The medium is a fluid only over scales exceeding the correlation length.

 It turns out that finiteness of the correlation length is incompatible with the vanishing of various transport coefficients like shear viscosity and heat conductivity, another feature of the ideal fluid paradigm.  For a gas there is a simple way to see this.  The usual estimates of the mentioned transport coefficients~\cite{Reichl} have them proportional to the mean free path of the gas' molecules.  But in a gas the mean free path and correlation length are the same thing.  Thus the finite correlation length forces the transport coefficients of the gas, in particular the shear viscosity, to be nonvanishing.  We shall see in the sequel that the same conclusion applies generally to any liquid as well.  Thus a fluid can be fully compatible with the GSL only if it is dissipative to some extent (and thus not ideal).  Another way of putting this is that a fluid with arbitrarily small shear viscosity is unphysical.

 \subsection{Lower bound on correlation length}

How large must the correlation length $l$ be?  Microscopically speaking it must obviously be at least as large as the intermolecular distance.  But can we say something without delving into the structure of the fluid?   Let us approach this question in the spirit of Wilson's work on the renormalization group~\cite{Wilson}. We consider
some thermodynamic system with the typical linear size $R$. Because the system is macroscopic (by definition), one can
represent any extensive thermodynamic variable as some quantity times the system's volume $V$. For example, one can write the entropy as $S=sV$ and the energy as $E=\rho V$.
Now decrease $R$. For sufficiently small $R$, already comparable with the system correlation length $l$, the system ceases to be macroscopic, and extensivity is generally lost.  This means that the expressions for the entropy and energy densities themselves become dependent on $R$.  The system is no longer a continuum.  Can one set a lower bound on the correlation radius $l$ purely from macroscopic considerations?

The standard answer to the above question would be in the negative. But, surprisingly, the true answer is yes. As we have seen, for a macroscopic system the UBE can be restated in the form~(\ref{Bekensteinforfluid}), which immediately leads to the inequality
\begin{equation} \label{Bekenstein11}
l \gtrsim \frac{s}{2\pi \rho}\,.
\end{equation}
The main point here is that a macroscopic system with size much smaller than $s/2\pi \rho$ would violate the UBE; thus the correlation length cannot be much smaller.  We emphasize, again, that $l$ may easily be much larger than the minimal scale~(\ref{Bekenstein11}). The above result answers the following question: given a macroscopic system with given entropy and energy densities,
what is its minimal possible correlation length? It shows that macroscopic quantities do ``know" about the minimal correlation length of the system.

\subsection{Lower bound on shear viscosity}
\label{lowervisc}

We mentioned earlier that a nonvanishing correlation length $l$ ensures that the transport coefficients of a gas (which are proportional to $l$)
do not vanish. For example, for shear viscosity one has the order of magnitude estimate~\cite{Reichl}
\begin{equation}
\label{estimate}
\eta \sim \rho l a,
\end{equation}
where the speed of sound $a$ is of the order of molecular speed~\cite{Landau10}.
The above estimate together with
Eq.~(\ref{Bekenstein11}) give that $\eta/s$ is subject to the inequality
\begin{equation}
\label{new1}
\frac{\eta}{s} \gtrsim \frac{a}{2\pi}\,.
\end{equation}

Let us show that the estimate~(\ref{estimate}), and thus the bound~(\ref{new1}), must hold also for liquids sufficiently far from a critical point. In this preliminary treatment we neglect heat conductivity and bulk viscosity. We note that, for any fluid, density perturbations at scale much larger than $l$ generate sound waves. On the other hand, perturbations with scale much smaller than $l$ are not coherent and produce no sound. Furthermore,
one can still use hydrodynamics asymptotically to describe the evolution of perturbations with scale $l$. Then the demand that there is no well-defined sound
at smaller scales gives the asymptotic condition that at scale $l$ the wave decay time $\sim \rho l^2/\eta$, as found from hydrodynamics~\cite{Landau6}, should be comparable with the wave period $\sim l/a$. This produces the estimate~(\ref{estimate}) for liquids.

Alternatively, the results~(\ref{estimate})-(\ref{new1}) can be recovered by considering a sound wave already propagating through the liquid.  Its Fourier components with reduced wavelengths near or below $l$ should decay over a distance comparable to $l$ since we cannot have macroscopic flow at those smaller scales.  Now from the \textit{macroscopic} point of view, the decay must be caused by transport processes controlled by the viscosities or heat conductivity. According to fluid theory~\cite{Landau6}, a sound wave of wavelength $\lambda$ penetrates a distance of order $\rho a (\lambda/2\pi)^2/\eta$ into the liquid before damping out.  According to the above argument, for $(\lambda/2\pi)\lesssim l$ this penetration length should be $l$, which gives Eq.~(\ref{estimate}) again.

For relativistic fluids the above consideration should be modified slightly. Here the wave decay time
includes $\rho+p$ instead of $\rho$ \cite{Weinberg}, which leads to
\begin{equation}\label{est-rel}
    \eta \sim (\rho + p) l a\ .
\end{equation}
For most realistic fluids $p\leq \rho/3$, see \cite{Landau5}, and $\rho+p\sim \rho$, so
this modification is not essential for the order of magnitude estimate of $\eta$, see
the next subsection for an example.

The above estimates, however, must fail sufficiently close to a critical point. Both $\eta$ and $l$ diverge at the critical temperature
with the power law $\eta\propto l^{\Delta}$ holding in the vicinity of the critical point. Here $\Delta$ is much less
than $1$~\cite{BFBS}.  This is incompatible with Eq.~(\ref{estimate}) because both $\rho$ and $a$ are finite at the critical point. The reason for the failure of the estimate is complications in the asymptotic matching procedure above, related to the difference of the critical exponents of the various quantities involved.  Another issue  is that near a critical point the static correlation length $l$ becomes different from the scale beyond which hydrodynamics applies.

\begin{figure}
\includegraphics[scale=0.90]{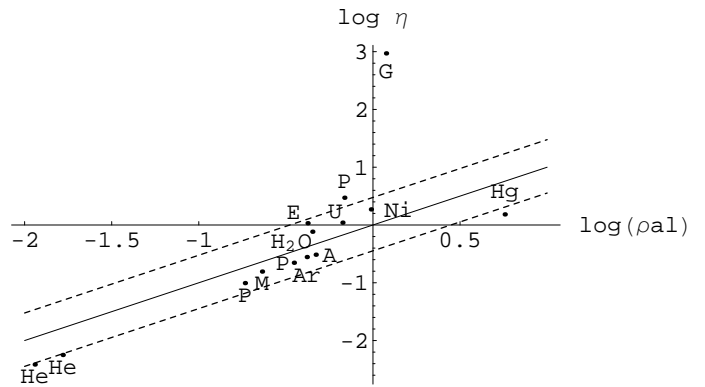}
\caption{Plot of log$_{10}$ of experimental viscosities of eleven pure liquids (data taken mostly from Ref.~\cite{CRC}) vs log$_{10}$ of the estimates from Eq.~(\ref{estimate}) with $l$ identified with the average intermolecular separation.  Both viscosities and estimates are in SI units (mPa s).  Nonstandard symbols are ``M'' for methane, ``P'' for propane, ``A'' for acetone, ``E'' for ethanol, ``U'' for undecane (C$_{11}$H$_{24}$), ``Ni'' for nitrobenzene (phenil-NO$_2$) and ``G'' for glycerol (C$_3$H$_8$O$_3$).  A repeated symbol correspond to viscosities at several pressures and temperatures.  The solid line is the locus of $\eta=\rho a l$; the dotted lines demarcate the region where the viscosity lies within a factor of 3 of $\rho a l$. }\label{fig:fig1}
\end{figure}

Nevertheless, sufficiently far from the critical
point the estimate~(\ref{estimate}) seems to produce remarkably good results, as shown by Fig.~\ref{fig:fig1} for a number of pure liquids.  (The only resounding failure is glycerol, one of the most viscous fluids known.  The problem with it may be that $l$ is much larger than the intermolecular separation, which we routinely used as estimator for $l$.)   This fact is remarkable because
it allows to reach conclusions on the magnitude of viscosity---quite possibly of practical value---even for liquids for which no microscopic theory is available.

Let us now turn to the lower bound on the viscosity-entropy density ratio~(\ref{new1}).   It evidently holds if the estimate~(\ref{estimate}) works.  Note that Eq.~(\ref{new1}) will most often be a strong inequality, particularly for nonrelativistic systems,
for which the UBE is known to be a very liberal bound.  Thus the bound~(\ref{new1}) is generally valid far from a critical point.  On the other hand, near the critical point the bound~(\ref{new1}) holds trivially because its l.h.s. diverges while its
r.h.s. remains finite. Thus the divergence of shear viscosity at the critical point~\cite{BFBS} only strengthens the bound.

The considerations in this subsection provide a strong argument in favor of the existence of a generic lower bound on the viscosity-entropy density ratio. The reformulation of the arguments with regard to the other transport coefficients,
including thermal conductivity, is left for future work.

\subsection{The quark-gluon plasma}

It was the challenge presented by the quark-gluon plasma (QGP) which motivated the activity leading to the formulation of the KSS bound.  As a practical use of our estimate~(\ref{estimate}), let us apply it to the QGP. Quantum chromodynamics (QCD) predicts the existence of such a form of matter where the usually confined quarks and gluons are essentially free forming a fireball with a typical size  $\sim 10$ fm. The transition from the hadronic state of matter to the QGP occurs at densities of $1-10$ GeV/fm$^3$. QCD lattice calculations predict that the transition should take place at a critical temperature $T_c \sim 160$ MeV, and that the speed of sound for a QGP above $T \approx 300$ MeV is about $a \sim 0.3 \,c$ (see Ref.~\cite{RHIC} and further references therein).

Measurements on the QGP at Brookhaven's relativistic heavy ion collider (RHIC) indicate that the  ratio of energy density $\epsilon$
to entropy density $s$ is roughly proportional to the QGP temperature $T$ (see, e.g., Fig. 3 in Ref.~\cite{Rafelski-Letessier}).  Specifically, for the high energy results of RHIC one has
\begin{equation}
\frac{\epsilon}{s} \simeq 0.8\,kT. \label{rhic}
\end{equation}
Using this and estimate~(\ref{estimate}) we have
\begin{equation}
\frac{k\,\eta}{s\,\hbar} \sim  \frac{k\,\epsilon\, a\, l}{s\,\hbar c^2} \simeq \frac{0.8\,k\, T\, a \, l}{\hbar c^2} \approx 0.2
\end{equation}
where in the last step we assumed $a \sim 0.3 \,c$ at $kT= 500$ MeV, $l \sim 0.3$ fm corresponding to a number density $n \sim 40$ /fm$^3$, and employed $\hbar c \approx 200$ MeV fm.

Our estimate for the QGP viscosity is thus in harmony with the KSS bound~(\ref{KSS}). Moreover, it is in quite reasonable agreement with the experimental data which require a shear viscosity to entropy density ratio as low as $\eta/s \leq 0.2$ \cite{Teaney,Lacey,DDGO,RR}.

\subsection{KSS bound---a tightened entropy bound for fluids}

We draw attention to the similarity between inequality~(\ref{new1}) with $c$ and $\hbar$ restored, namely
\begin{equation}
\label{new2}
\frac{\eta}{s} \gtrsim \frac{a/c}{2\pi}\hbar\,,
\end{equation}
and the conjectured KSS bound~(\ref{KSS}).  At first sight this inequality seems to fall well below the KSS bound since in many cases $a\ll c$.  However, we must keep in mind that, especially in such nonrelativistic circumstances, we expect the inequality to be a strong one.  The minimum $\eta/s$ may thus be well above the literal bound~(\ref{new2}), and not necessarily at variance with Eq.~(\ref{KSS}). For relativistic media $a\sim c$, and there is no difference to speak of between bounds~(\ref{KSS}) and~(\ref{new2}).  Thus arguments based on the UBE, and ultimately on the GSL, seem to suggest a \textit{raison d'\,\^{e}tre} for the mysterious KSS bound.

Before proceeding we wish to provide an alternative viewpoint. The KSS bound Eq.~(\ref{KSS}) is usually interpreted as saying that a fluid cannot be
too perfect.  However, in the form
\begin{equation}
\label{KSSentropy}
s\leq 4\pi \eta,
\end{equation}
the bound is really an entropy bound, specifically an upper bound on the entropy density of an arbitrary fluid (we return to the use of units with $c=\hbar=1$ as in the previous sections). Clearly the above bound is reminiscent
of the UBE, Eq.~(\ref{Bekensteinentropy}), and its fluids version,  Eq.~(\ref{Bekensteinforfluid}).

But at least for non-relativistic fluids, the KSS bound is a tighter entropy bound by orders of magnitude than the UBE.  Restoring the speed of light $c$ we introduce the variable $\eta_0= l c \rho$ ($l$ being, again, the correlation length, either the mean free path for a gas, or, typically, the intermolecular separation for a liquid) which has the dimensions of viscosity.  Thus Eq.~(\ref{Bekensteinforfluid}) gives $s< 2\pi \eta_0$.
Using the estimate~(\ref{estimate}) we have $\eta/\eta_0\sim a/c$. It follows that for nonrelativistic fluids Eq.~(\ref{KSSentropy}) is indeed a much tighter entropy bound than Eq.~(\ref{Bekensteinforfluid}).    Thus the KSS bound can be viewed as a tightened version of the UBE in  guise of Eq.~(\ref{Bekensteinforfluid}) which is available for the class of systems exhibiting macroscopic fluid behavior.

Both the ideal fluid paradox of Sec.~\ref{paradox} and the interpretation of the KSS bound as an entropy bound for fluids suggest the GSL as a natural frame for investigating relations between entropy bounds.  Because the UBE is obtained in the most transparent way via the GSL, one may hope that the KSS bound should likewise be obtainable  most simply with help of the GSL.   We turn now to considerations in this direction.

\subsection{Where does the KSS bound come from?}

First we remark that it is not clear how to obtain the KSS bound directly from microscopic physics. Inspection  of the Green-Kubo formula~\cite{Landau9} for the viscosity in terms of fluctuations shows no apparent connection of the viscosity and the entropy.  Such consideration affords no special status to  the ratio $\eta/s$, at least not from the viewpoint of  nongravitational physics. On the other hand, the KSS conjecture emerged from holographic type arguments that connect quantum field theory with gravity. It might thus turn out that a derivation of the KSS bound requires use of the still nonexistent theory of quantum gravity.   But even if this were true, one need not loose heart.   There is general agreement that black hole entropy, and the GSL which hinges on it, reflect some aspect of quantum gravity.

The KSS is not universally accepted.  Many objections to it rely on scenarios where very many particle species  are supposed to exist in nature (see Ref.~\cite{ChermanCohen} for an example and references).  These are the same objections raised against the validity of the UBE (see Ref.~\cite{JB4} for references).  In fact, any entropy bound whatsoever invites attacks of this sort, for if the number of species that may show up is unlimited, the entropy can be made as large as desired while keeping parameters like energy or total particle number  fixed.  Although it is formally true that  with many species available, the KSS bound must fail, this does not detract from its heuristic usefulness; one is often interested on the entropy of a specific system, one whose particle content is fixed ahead of time.   It should also be mentioned that many particle scenarios, if not arbitrarily legislating particle proliferation, conjure up fine-tuned or baroque setups to beget the required large number of species of quasiparticles or excitations.  This aspect considerably decreases the appeal of the species proliferation arguments.

Both of the above considerations make it clear that a ground up approach to deriving the KSS bound is unlikely to succeed.  Alternative, indirect approaches are needed.  A point in favor of employing the GSL to investigate the origin of  the  KSS bound is that the GSL ``knows" the actual number of species in nature~\cite{JB4,JB5}.  For example, black hole entropy, which plays a crucial role in the GSL, should in principle depend on the number of elementary fields, yet all derivations endow it with a fixed coefficient which may be thought as determined by the actual number of species.   This feature would act to neutralize the above mentioned argument against the validity of the KSS bounds.

\section{Summary and outlook}
\label{summary}

We have worked out the expression for the rate of change of the total entropy of a system consisting of a Schwarzschild
black hole and ideal fluid which can accrete onto the hole. The well known Bondi flow is the case of fast accretion with velocity near to the speed of
light; for it the GSL is always obeyed due to the UBE.
For sufficiently slow accretion velocity the flow violates
the GSL.   The question of whether flow with the required small accretion velocity is a realistic option is answered in the affirmative by a critical revision, in the Appendix, of the venerable argument~\cite{Gibbons1} claiming that it is impossible to adiabatically lower mass to near the black hole horizon.

 Since it is known from microscopic considerations~\cite{Sorkin}  that any physical system should comply with the GSL, the above result would constitute a serious paradox if  the ideal fluid is a continuum, as usually considered.  We find, however, that the paradox can be defused if one takes into account that the continuum picture must break down at some level by virtue of the fluid having a nonvanishing correlation length.  An auxiliary role in the nullification of the paradox is played by the universal entropy bound.  It gives a lower bound on the correlation length in relation to the fluid's entropy density.

The lower bound on the correlation length of an arbitrary thermodynamic system in terms of its (macroscopic) entropy and energy densities is an unexpected and thought-provoking aspect of the UBE. It makes it clear that the GSL, though a macroscopic law, ``knows'' about the microscopic structure of matter.  The lower bound must rank as one of the most impressive consequences of the extension of the second law of thermodynamics to include black holes.

The breakdown of the continuum description of a fluid has momentous consequences: discreteness of matter and thermal fluctuations necessarily engender nonideal behavior parameterized by the viscosity and heat conduction coefficients.  We have shown that one can expect the existence of a universal lower bound on the viscosity of an arbitrary fluid. Our argument yields a seemingly novel estimate for the viscosity of a fluid far from the critical region; this estimate is shown to be better than an order of magnitude estimate for a range of pure liquids. In addition, the estimate
works reasonably well for the quark-gluon plasma, which viscosity was a subject of much study lately \cite{Teaney,Lacey,DDGO,RR}.

Together with the lower bound on the correlation length, the viscosity estimate yields a lower bound on viscosity per unit entropy density which may be expressed in terms of the sound velocity of the fluid. The appearance of the speed of sound here, in contrast to the (implicit) speed of light in the original KSS arguments, may signify that the KSS argument can be improved, or may perhaps disclose that the field theories considered by KSS as representatives of strongly coupled fluids are not sufficiently generic---in all these theories the sound velocity is comparable  with the speed of light. An additional issue is whether the bound can be extended to other transport coefficients.

If there is indeed a lower bound on the ratio of viscosity to entropy density for an arbitrary fluid, then it is likely to be rooted in some very basic  principle of physics. We have shown that the generalized second law of thermodynamics, combining as it does gravity with thermodynamics, can be  such a principle. Starting from it, one can anticipate that the viscosity to entropy ratio of an arbitrary fluid obeys a certain lower bound.  In particular, the GSL incorporates physics that weakens the recent objections~\cite{ChermanCohen} to the bound's very existence.  On the basis of the above, we propose that the use of the GSL may lead to a clarification of the KSS bound's origin. Although we have made no concrete progress towards reaching a final sharp inequality, we have given for the first time physical arguments indicating that this elusive bound, whose existence is sometimes disputed, indeed exists.

\acknowledgments

I. F. thanks V. Lebedev for an illuminating discussion.  G. B. thanks the Hebrew University for a Golda Meir fellowship.  J. D. B.'s research is supported by grant 694/04 of the Israel Science Foundation, established by the Israel Academy of Sciences.

\appendix

\section{Equilibrium of matter near a black hole and realizability of slow accretion flows}

Gibbons~\cite{Gibbons1} claimed that one cannot lower a rope to near a black hole at an arbitrarily small
velocity, without it being torn at some finite distance from the horizon. The conclusion was based
upon the assertion that the stresses that arise in the rope become so strong that they violate the weak energy condition \cite{HP,Hawking-Ellis} which must be obeyed by the energy-momentum tensor of physical matter.  Here we reexamine the problem
and show that, in fact, the energy condition need not be violated and thus that adiabatic lowering down to the horizon is, in principle, possible.

In the stationary state of some matter in the vicinity of a spherical black hole, the energy momentum
tensor obeys
\begin{equation}
0=T^{\nu}_{\mu; \nu}=\frac{1}{\sqrt{-g}}\frac{\partial (T^{\nu}_{\mu}\sqrt{-g})}{\partial x^{\nu}}-\frac{1}{2} \frac{\partial g_{\nu\rho}}{\partial x^{\mu}}\,T^{\nu\rho}\ ,\label{div}
\end{equation}
where we used the symmetry of $T_{\mu\nu}$~\cite{Landau2}.  Below we concentrate on  configurations for which
$T^{\mu}_{\nu}$ is diagonal. Using the same coordinates as in Sec.~\ref{balance}, the $\mu=0$ component of the above equations is satisfied automatically,
while the radial component gives
\begin{eqnarray}&&\label{radial}
\frac{1}{r^2}\frac{\partial \left(T^r_rr^2\right)}{\partial r} - \frac{MT^0_0}{r^2(1-2M/r)}
+ \frac{MT^r_r}{r^2(1-2M/r)}\nonumber\\&&-\frac{T^{\theta}_{\theta}+T^{\phi}_{\phi}}{r}=0.
\end{eqnarray}

The above equation gives a linear relation between the different components of the stress tensor which need not, \textit{a priori}, respect the weak energy condition. As an example, consider a thin spherical shell in equilibrium.
For the shell the contribution of $T^r_r$ in the above equation is negligible. This can be seen by noting that $T^r_r$
vanishes at the shell boundaries, and as a result its values within the shell vanish together with the ratio of shell thickness to radius.   Far from the hole, where the shell is describable by the classical
linear elasticity theory, the maximal value of $T^r_r$ is found to be proportional to the square of the above ratio.  So the thin shell is basically supported by the tangential stresses  $T^{\theta}_{\theta}$ and $T^{\phi}_{\phi}$, which must be equal because of spherical symmetry.   Eq.~(\ref{radial}) gives $|T^{\theta}_{\theta}|/|T^{0}_{0}|=M/(2r-4M)$. Now the weak energy condition would demand $|T^{\theta}_{\theta}|/|T^{0}_{0}|<1$.  It is thus clear that a stationary thin shell would violate the
condition at $r<5M/2$. Thus a physical
thin shell, i.e. one obeying the energy condition,  cannot support itself arbitrarily close to the horizon (see Ref.~\cite{FHK} for the more detailed discussion).

Ref.~\cite{Gibbons1} argued that the rope cannot be in equilibrium with its lower end arbitrarily close to the black hole, similarly to the thin shell above. Here we wish to correct this conclusion. We shall assume that the rope fibers can be considered radial, such as in the case of a conical rope filling the portion of space defined by some solid angle. Since the rope can only
support stresses along its fibers,  the general form of its energy-momentum tensor  is $T^{\mu}_{~\nu}= \mathrm{diag}[-\rho,S,0,0]$, and Eq.~(\ref{radial})
gives
\begin{equation}
 \label{drS}
\frac{\partial \left(r^2S\right)}{\partial r} = -\frac{M(\rho+S)}{1-2M/r}\,.
\end{equation}

The above linear equation can be used to express the stress $S(r)$ in terms of the rope density $\rho$
and the boundary condition at the lower end of the rope, $r=r_0$, which is defined by the load.
The solution can be written as a sum $S(r)=S_1(r)+S_2(r)$, where $S_1(r)$ describes the stresses
caused by the load, while $S_2(r)$ describes the stresses caused by the rope's own weight. The expression
for $S_1(r)$ (the solution for a weightless rope) expresses the ``constancy of the tension'' along the rope:
\begin{equation}
S_1(r)=\frac{r_0^2\sqrt{1-2M/r_0}}{r^2\sqrt{1-2M/r}}\,S(r_0),
\end{equation}
where  $S(r_0)$ is determined by the weight of the load; for a point mass $m$ we have $r_0^2\sqrt{1-2M/r_0}\,S(r_0)=-m$.   Note that $S_1(r)$ is negative, as befitting a tension.  The ``constancy of the tension'' gives monotonically decreasing $|S_1(r)|$ as one moves out because the rope cross-section increases with $r$ by the assumed symmetry of the fibers.

The expression
for $S_2(r)$ (the solution without the load) describes how the force caused by the rope above some $r$ balances
the weight of the rope below:
\begin{equation}
\label{a1}
S_2(r)=-\frac{1}{r^2\sqrt{1-2M/r}}\int_{r_0}^r dr'\frac{M\rho(r')}{\sqrt{1-2M/r'}}\,.
\end{equation}
For definiteness we shall consider below the case of the rope with a constant density $\rho_0$ where integration gives
\begin{eqnarray}&&
S_2(r) = -\frac{M\rho_0}{r^2\sqrt{1-2M/r}}\Biggl[ r\sqrt{1-2M/r} \\&& -  r_0\sqrt{1-2M/r_0}
 + M\ln  \frac{r-M+r\sqrt {1-2M/r}}{r_0-M+r_0\sqrt {1-2M/r_0}} \Biggr]. \nonumber
\end{eqnarray}

We now show that the above solution does not violate the weak energy condition $|S|/\rho_0\leq 1$, and in particular, that
no violation occurs even when the rope's end is arbitrarily close to the horizon. By
choosing sufficiently large $\rho_0$ we may always disregard $S_1(r)$, so it is sufficient to show that
$\max|S_2(r)|/\rho_0\leq 1$. It is easy to see from Eq.~(\ref{a1}) that the maximum of $|S_2(r)|=-S_2(r)$,
at given $\rho_0$ and $M$, grows as $r_0$ decreases. This just means that the maximal stress is the bigger the closer
the lower end of the rope is from the horizon. Therefore, it is enough to show that $\max|S_2(r)|/\rho_0\leq 1$ for $r_0$ which is infinitesimally close to $r_H=2M$. In general, $|S_2(r)|$, vanishing as it is at
both endpoints, $r=r_0$ and $r=\infty$, has a unique maximum at $r_*\in(r_0,\infty)$.  It is easy to see
numerically that in the limit $r_0\to r_H$ the point $r_*$ also tends to $r_H$. Defining $\epsilon\equiv (1-2M/r_*)^{1/2}$ we find analytically
\begin{eqnarray}&&
\sqrt{1-\frac{2M}{r_0}}=\frac{8\epsilon^3}{3}+\ldots
,\nonumber \\&&
\frac{\max|S_2(r)|}{\rho_0}\approx 1-4\epsilon^2+\ldots
,
\end{eqnarray}
where $\epsilon\ll 1$ and the dots stand for subleading terms.
It follows from the above that the weak energy condition is obeyed for any $r_0>r_H$.
Thus, at least from the viewpoint of the energy conditions, it is possible in principle to lower  a body adiabatically all the way down to the horizon by means of a suitably constructed rope .

Our result is at variance with that of Gibbons~\cite{Gibbons1}.
The discrepancy arises because he missed a term when calculating a certain 4-divergence (Eq.~(5) in Ref.~\cite{Gibbons1}). By taking this term into account, the main result, Eq.~(8) in Ref.~\cite{Gibbons1}, now reads
\begin{eqnarray}&&
- \frac{\mathrm{d}T}{T+\sigma} = \frac{\mathrm{d} V}{V}\,. \label{Gibb}
\end{eqnarray}
Here $T=SA$ is the tension, $A$ the rope's cross section, $\sigma=\rho A$ the energy per unit proper length of the rope, and $V=\sqrt{-K_a\,K^a}$ is the norm of the timelike Killing field $K^a$ (the spacetime is taken to be stationary). In the original derivation, the term $T$ in the denominator of the l.h.s. of Eq.~(\ref{Gibb}) is absent. The corrected Eq.~(\ref{Gibb}) is indeed  in accord with our Eq.~(\ref{drS}). Observing that in the Schwarzschild geometry  $V=\sqrt{1-2M/r}$ and further $A^{-1}\mathrm{d}A/\mathrm{d}r = 2/r$ for the case of a conical rope, one easily checks that Eq.~(\ref{Gibb}) reduces to our Eq.~(\ref{drS}).

Our analysis demonstrates that no problem of principle militates against the creation of the boundary
conditions required for the slow accretion flow considered in Sec.~\ref{paradox} here.  A suitably designed rope could be used to deposit parcels of fluid at rest near the horizon.



\end{document}